# AlGaN /GaN superlattice based p-channel field effect transistor (pFET) with TMAH treatment

*Athith Krishna[1,*], Aditya Raj[1], Nirupam Hatui[1], Onur Koksaldi[1], Raina Jang[1], Stacia Keller[1], and Umesh K Mishra[1]*

Athith Krishna, Aditya Raj, Dr. Nirupam Hatui, Dr. Onur Koksaldi, Raina Jang, Dr. Stacia Keller and Prof. Umesh K. Mishra
Department of Electrical & Computer Engineering, University of California – Santa Barbara, Goleta, California – 93117. United States of America
E-mail: athith@ucsb.edu

To realize the full spectrum of advantages that the III-nitride materials system offers, the demonstration of p-channel III-nitride based devices is valuable. Authors report the first p-type field effect transistor (pFET) based on an AlGaN/GaN superlattice (SL), grown using MOCVD. Magnesium was used as the p-type dopant. A sheet resistance of 11.6 k$\Omega$/■, and a contact resistance of 14.9 $\Omega$.mm was determined using transmission line measurements (TLM) for a Mg doping of $1.5 \times 10^{19}$ cm$^{-3}$ of Mg. Mobilities in the range of 7−10 cm$^2$/Vs and a total sheet charge density in the range of $1 \times 10^{13} – 6 \times 10^{13}$ cm$^{-2}$ were measured using room temperature Hall effect measurements. Without Tetramethylammonium hydroxide (TMAH) treatment, the fabricated pFETs had a maximum drain-source current ($I_{DS}$) of 3mA/mm and an On-Resistance ($R_{ON}$) of 3.48 k$\Omega$.mm, and did not turn-off completely. With TMAH treatment during fabrication, a maximum $I_{DS}$ of 4.5mA/mm, $R_{ON}$ of 2.2k$\Omega$.mm, and five orders of current modulation was demonstrated, which is the highest achieved for a p-type transistor based on (Al,Ga)N.

## 1. Introduction

Gallium nitride (GaN) based solid-state lighting has revolutionized the optoelectronics industry due to its efficiency, low cost and durability.[1] GaN, other III-nitrides and their alloys are the preferred material system for next-generation electronic devices, realizing the needs of advanced communication systems, power conversion and energy conservation, thus enabling compact and affordable electronic systems.[2-10] The wide-bandgap of GaN (~3.4eV) enables devices with higher breakdown voltages, and the built-in polarization fields in AlGaN/GaN heterostructures lead to a very high mobility and charge in two-dimensional electron-gases (2DEG).[11,12] But, to tap into the versatility of this materials system, complementary MOS (CMOS) circuits based on GaN and its alloys are desired. CMOS based on GaN has the potential to adopt Si based CMOS topologies, and can be used to realize very efficient gate drivers and high-voltage DC-to-DC converters.[13] Unfortunately, like in most wide-bandgap materials, the properties of holes in GaN has limited the implementation of p-channel devices in the past.

The high effective mass of holes in GaN results in a low hole mobility, with the maximum observed value being 40 cm$^2$/Vs at a hole concentration of $2 \times 10^{12}$ cm$^{-2}$ in a two dimensional-hole gas (2DHG), and 20 cm$^2$/Vs in bulk p-GaN (p=$1 \times 10^{17}$ cm$^{-3}$).[14,15] In addition, Mg, which is commonly used as the p-type dopant is a deep acceptor in GaN and AlGaN (160 - 220 meV).[16,17] Furthermore metal-organic chemical vapor depostion (MOCVD) grown p-GaN material, like the one reported here, is passivated by hydrogen in the as-grown state and needs annealing at a high temperature (here, 825$^0$C) for activation.[18] Additionally, the high p-GaN work function at typical doping levels results in challenges to making ohmic contacts to p-type GaN.[19]

In view of these challenges, III-nitride based p-FETs have received significantly less attention compared to GaN-based HEMTs.[14, 20-27] In this report, we present the first p-type GaN/AlGaN superlattice pFET. Based on a reported AlGaN/GaN p-MOS, the On-Resistance ($R_{ON}$) of the device can be modelled according to **Equation (1)**.[25]

$R_{ON} = (2 \times R_{contact}) + R_{channel} + R_{access-source} + R_{access-drain}$ (1)

Depending on the design of the device structure, the use of GaN/AlGaN superlattice(SL) can help reduce either the $R_{access}$ or $R_{channel}$ of the device. Application of periodic oscillations to the valence band edge can help to increase the doping efficiency of acceptors in p-type III-nitride materials.[28] With polarization effects aiding the process, SLs lead to the necessary valence band edge oscillations.[29] As demonstrated by Kozodoy et al.,the use of AlGaN/GaN SLs lead to an increased overall hole concentration as it facilitates the ionization of deep acceptors in the barriers into the valence band of the narrower bandgap material (here, GaN).[30] SLs also give rise to 2DHG formation at the GaN/AlGaN interfaces, thus leading to improved mobility values in the channel, and reduced resistances.

**2. Ga-polar Uniformly doped p-type Al$_{0.2}$Ga$_{0.8}$N/GaN SL – Material growth**

**Figure 1** shows the p-(GaN/AlGaN) SL epitaxial structure. MOCVD technique was used to grow the epitaxial samples on sapphire substrates. First, 1.5µm of semi-insulating GaN was deposited, followed by 16nm of uniformly doped p-type GaN:Mg with [Mg]=$4.5 \times 10^{19}$ cm$^{-3}$. This layer was used to counter dope polarization related 2DEG formation during Al$_{(x)}$Ga$_{1-x}$N deposition. Above this, 20nm Al$_{(x)}$Ga$_{1-x}$N:Mg was grown where the composition was graded from x=0 – 12% followed by 4nm Al$_{0.2}$Ga$_{0.8}$N to be used as the back-barrier for hole transport. The next layers consisted of the SL stack with 4,7 and 10 SL periods of 8nm p-GaN/8nm p-Al$_{0.2}$Ga$_{0.8}$N uniformly doped with $1.5 \times 10^{19}$ cm$^{-3}$ of Mg. Finally, a 20nm p-GaN cap was grown, which was doped with $4.5 \times 10^{19}$ cm$^{-3}$ of Mg, to act as the contact layer. The as-grown material was characterized using secondary ion mass spectroscopy (SIMS), X-ray diffraction (XRD) and atomic force microscopy (AFM), (not shown). The electrical properties of the planar samples were investigated by room temperature Hall and transmission line measurements (TLMs). The latter were carried out with device isolation. Reported p-FET devices were fabricated from 4 period SL samples.

Since the Mg acceptors were passivated in the as-grown material, optimization of the post growth activation annealing was carried out as well. 100nm SiO$_2$ was deposited using PECVD on the as-grown samples to protect their surface. 3 minute rapid thermal annealing (RTA) was carried out for Mg dopant activation at temperatures in the range of 675$^0$C-900$^0$C in nitrogen and nitrogen/oxygen ambient. **Figure 2** shows the sheet resistance as a function of activation temperature. The optimum activation temperature was found to be 825$^0$C and was used for all the samples processed.

**Figure 3** shows total hole sheet charge density (p$_s$) and hole mobility as a function of the Mg doping in samples with 10 SL periods. The p$_s$ decreased from $6.25 \times 10^{13}$ cm$^{-2}$ to $1 \times 10^{13}$ cm$^{-2}$ as the Mg doping was increased from $1.5 \times 10^{19}$ cm$^{-3}$ to $6 \times 10^{19}$ cm$^{-3}$. This behavior may be due to the self-compensation and/or formation of Mg clusters at high Mg doping.[31] Over the same doping range, the hole mobility, μ$_p$, increased from 6 cm$^2$/Vs to 10.2 cm$^2$/Vs, increasing with decreasinghole sheet charge density. **Figure 4** displays the sheet resistance of the same samples determined in the TLM measurements. The data confirm the observation of the Hall measurements, and the sheet resistance increased as the Mg doping was raised. The lowest sheet resistance of 11.6kΩ/■ was obtained for the samples with the lowest Mg doping of $1.5 \times 10^{19}$ cm$^{-3}$ in the SL.

**Figure 5** shows the total sheet charge density and the hole mobility as a function of the SL period for samples with $1.5 \times 10^{19}$ cm$^{-3}$ Mg doping. p$_s$ increased as the SL period was increased from 4 to 7, but slightly decreased again when the SL period was increased from 7 to 10 SL periods. Most likely, not all the SL periods in the 10-period sample were contacted by the top p-contact, leading to an effectively lower charge than expected.The hole mobility was largely invariant as the SL periods were varied from 4 to 10.

## 3. Planar single-channel pFET process flow

For device fabrication, a 4 period SL sample with a Mg doping of $1.5 \times 10^{19}$ cm$^{-3}$ was chosen. The design used for device fabrication was a 3-period SL gate recess, as shown in **Figure 6**. **Figure 7** illustrates the process flow for the devices, without TMAH treatment. The fabrication steps were as follows: (1) PECVD SiO$_2$ was deposited on the as-grown sample followed by 3-minute RTA activation at 825$^O$C; (2) SiO$_2$ was removed followed by ohmic metal deposition for source and drain. A palladium (20nm)/ gold (200nm) metal stack was used for the ohmic contacts; (3) A low power etch was carried out for the mesa etch to isolate the devices; (4) A low power etch was also used for the 3-period SL gate recess etch to leave one conductive channel under the gate; (5) Atomic layer deposition (ALD) technique was used to deposit 10nm of aluminum oxide ($Al_2O_3$), to be used as the gate dielectric; (6) A gate metal stack of titanium/gold was deposited. The fabricated device had a gate length of 0.5µm, a gate-source spacing of 1.0µm, and a gate-drain spacing of 2.0µm. For devices fabricated with TMAH treatment, the treatment as performed at $80^0C$ for 30 min after gate etch (step 4 in Figure 7), to remove any material which was damaged during the gate dry etch, from the sidewalls.

Plasma etch damage of GaN under gate and sidewalls can lead to gate leakage hampering the current modulation by the gate. It has been reported that TMAH treatment at elevated temperature removes the damaged region from the sidewall and makes it smooth and vertical.[32] **Figure 8** shows the SEM images before and after TMAH treatment. For the device with TMAH treament, 10 nm $SiO_2$ was deposited by ALD to serve as dielectric. The fabricated device had a gate length of 0.6µm, a gate-source spacing of 0.4µm, and a gate-drain spacing of 0.7µm.

## 4. Device characterization

### 4.1. Devices without TMAH treatment

**Figure 9** shows the output characteristics of pFET devices fabricated without TMAH treatment. The device exhibited a maximum drain-source current ($I_{DS}$) of 3mA/mm and an On-Resistance ($R_{ON}$) of 3.48kΩ.mm. The device did not pinch-off due to gate leakage and $R_{ON}$ was dominated by contact resistance and valence band discontinuities for vertical transport.

### 4.2. Devices with TMAH treatment

As described before, the TMAH treatment was implemented after the gate etch to remove any damaged material and hence reduce the gate leakage and subsequently pinch-off the device. Using TLM measurements, a sheet resistance($R_{sh}$) of 11.6 kΩ/■ and a contact resistance($R_c$) of 14.9Ω-mm were obtained, as shown in **Figure 10**. The fabricated transistors exhibited a maximum current of 4.5 mA/mm for $V_{GS}$ of -9 V and $V_{DS}$ of -10V (**Figure 11**) . The $I_{DS}$ as a function of $V_{DS}$ plot showed a clear current saturation and effective gate control, while $I_{DS}$ as a function of $V_{GS}$ curve showed five orders of modulation - $I_{on}/I_{off}$ ~$10^5$ (**Figure 12**). Linear scale transfer characteristics showed a pinch-off voltage of 2V and $g_{max}$ of 0.35 mS/mm (**Figure 13**). The output charcteristics also depicted a slight turn-on voltage, which was due to the non-ohmic nature of metal contacts.

**5. Conclusion**

With a maximum $I_{DS}$ = 4.5mA/mm, the AlGaN/GaN SL devices with TMAH etch during fabrication demonstrated the highest current amongst reported p-channel FETs based on AlGaN/GaN-only materials system, to date. Improvements in the device performance are expected through - further optimization of (a) contacts and gate dielectric, and (b) the composition, number of periods, and doping in the p-(GaN/AlGaN) SL. This report presents the first demonstration of an AlGaN/GaN SL based p-FET which along with already existing GaN nFETs can enable GaN based wide-bandgap CMOS. To achieve the complete benefit of using the superlattice, a FinFET or tri-gate device structure is to be fabricated to contact all the active p-type channels in the device. Wide-bandgap CMOS can open the path to achieving GaN based integrated circuit systems, and efficient and fast single-stage 48V to 1V DC-DC converters.

**Acknowledgements**
This work was supported by ASCENT, one of six centers in JUMP, a Semiconductor Research Corporation (SRC) program sponsored by DARPA.

Received: ((will be filled in by the editorial staff))
Revised: ((will be filled in by the editorial staff))
Published online: ((will be filled in by the editorial staff))

References

[1] S. Nakamura, S. Pearton, G. Fasol, *The Blue Laser Diode*. Springer Berlin Heidelberg, **2000**.

[2] S. Chowdhury, B.L. Swenson, M.H. Wong, U.K. Mishra, *Semiconductor Science and Technology* **2013,** *28*, 074014.

[3] M. Mußer, H. Walcher, T. Maier, R. Quay, M. Dammann, M. Mikulla, O. Ambacher, presented at The 5th European Microwave Integrated Circuits Conference, Paris, France, **2010.**

[4] A. Petersen, D. Stone, M. Foster, *Electronics Letters* **2017**, *53*, 1487.

[5] I. Chatzakis, A. Krishna, J. Culbertson, N. Sharac, A. J. Giles, M. G. Spencer, J. D. Caldwell, *Optics Letters* **2018,** *43*, 2177.

[6] R. Kemerley, H. Wallace, M. Yoder, *Proceedings of the IEEE* **2002,** *90*, 1059.

[7] U. Mishra, Y.-F. Wu, B. Keller, S. Keller, S. Denbaars, *IEEE Transactions on Microwave Theory and Techniques* **1998**, *46,* 756.

[8] P.J. Wellmann, *Zeitschrift Für Anorganische Und Allgemeine Chemie* **2017**, *643,* 1312.

[9] T.J. Flack, B.N. Pushpakaran, S.B. Bayne, *Journal of Electronic Materials* **2016,** *45,* 2673.

[10] N. Kolias, C. Whelan, T. Kaziorm, K. Smith presented at *IEEE MTT-S International Microwave Symposium*, **2010**.

[11] R. Quay, *Gallium Nitride Electronics*. Springer, Berlin, **2010**.

[12] J.P. Ibbetson, P.T. Fini, K.D. Ness, S.P. Denbaars, J.S. Speck, U.K. Mishra, *Applied Physics Letters,* **2000,** *77,* 250.

[13] S. Inoue, H. Akagi, *IEEE Transactions on Power Electronics*, **2002**, vol. 22, no. 2, pp. 535-542.

[14] B. Reuters, H. Hahn, A. Pooth, B. Holländer, U. Breuer, M. Heuken, H. Kalisch, A. Vescan, *Journal of Physics D: Applied Physics* **2014,** *47*, 175103.

[15] Y. Arakawa, K. Ueno, A. Kobayashi, J. Ohta, H. Fujioka, *APL Materials* **2016**, *4*, 086103.

[16] H. Amano, *Journal of The Electrochemical Society* **1990,** *137*, 1639 (1990).

[17] A.K. Viswanath, E.-J. Shin, J.I. Lee, S. Yu, D. Kim, B. Kim, Y. Choi, C.-H. Hong, *Journal of Applied Physics* **1998**, *83,* 2272.

[18] S. Nakamura, N. Iwasa, M. Senoh, T. Mukai, *Japanese Journal of Applied Physics* **1992,** *31*, 1258.

[19] J. Chen, W. Brewer, *Advanced Electronic Materials* **2015**, vol. 1, no. 8, p. 1500113.

[20] M. Shatalov, G. Simin, J. Zhang, V. Adivarahan, A. Koudymov, R. Pachipulusu, M.A. Khan, *IEEE Electron Device Letters* **2002,** *23*, 452.

[21] T. Zimmermann, M. Neuburger, M. Kunze, I. Daumiller, A. Denisenko, A. Dadgar, A. Krost, E. Kohn, *IEEE Electron Device Letters* **2004,** *25,* 450.

[22] G. Li, R. Wang, B. Song, J. Verma, Y. Cao, S. Ganguly, A. Verma, J. Guo, H.G. Xing, D. Jena, *IEEE Electron Device Letters* **2013,** *34,* 852.

[23] H. Hahn, B. Reuters, A. Pooth, A. Noculak, H. Kalisch, A. Vescan, *Japanese Journal of Applied Physics* **2013,** *52,* 128001.

[24] H. Hahn, B. Reuters, A. Pooth, H. Kalisch, A. Vescan, *Semiconductor Science and Technology* **2014,** *29,* 075002.

[25] R. Chu, Y. Cao, M. Chen, R. Li, D. Zehnder, *IEEE Electron Device Letters* **2016,** *37,* 269.

[26] S.J. Bader, R. Chaudhuri, K. Nomoto, A. Hickman, Z. Chen, H.W. Then, D.A. Muller, H.G. Xing, D. Jena, *IEEE Electron Device Letters* **2018,** *39,* 1848.

[27] N. Chowdhury, J. Lemettinen, Q. Xie, Y. Zhang, N.S. Rajput, P. Xiang, K. Cheng, S. Suihkonen, H.W. Then, T. Palacios, *IEEE Electron Device Letters* **2019,** *40,* 1036.

[28] E. Schubert, W. Grieshaber, I. Goepfert, *Applied Physics Letters*, **1996,** vol. 69, no. 24, pp. 3737-3739.

[29] P. Kozodoy, Y.P. Smorchkova, M. Hansen, H. Xing, S.P. Denbaars, U.K. Mishra, A.W. Saxler, R. Perrin, W.C. Mitchel, *Applied Physics Letters* **1999,** *75,* 2444.

[30] P. Kozodoy, M. Hansen, S.P. Denbaars, U.K. Mishra, *Applied Physics Letters* **1999,** *74,* 3681.

[31] N.A. Fichtenbaum, C. Schaake, T.E. Mates, C. Cobb, S. Keller, S.P. Denbaars, U.K. Mishra, *Applied Physics Letters* **2007,** *91,* 172105.

[32] M. Kodama, M. Sugimoto, E. Hayashi, N. Soejima, O. Ishiguro, M. Kanechika, K. Itoh, H. Ueda, T. Uesugi, T. Kachi, *Applied Physics Express* **2008,** *1,* 021104.

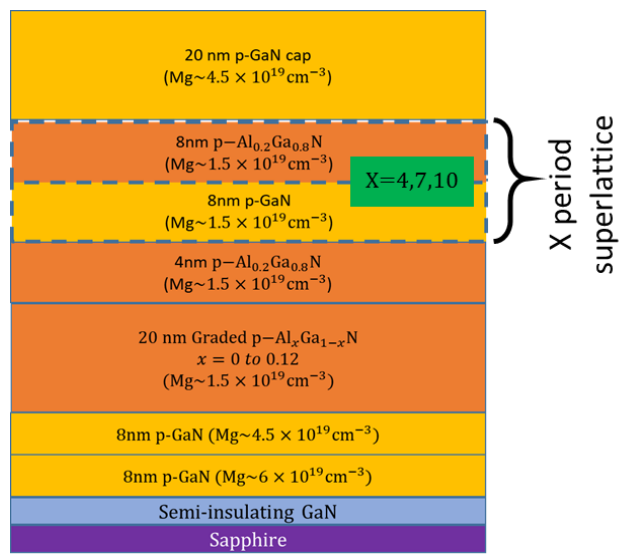

**Figure 1.** Design of the epitaxial structure for the superlattice p-(GaN/AlGaN) FET. Superlattice series with 4,7, and 10 SL periods was grown.

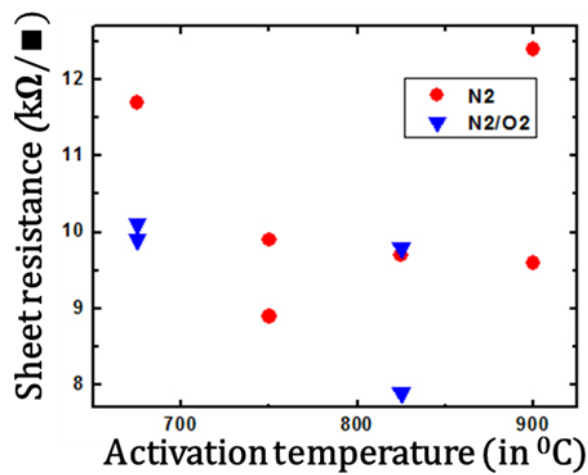

**Figure 2.** Sheet resistance as a function of activation temperature from the TLM measurements on non-isolated device samples.

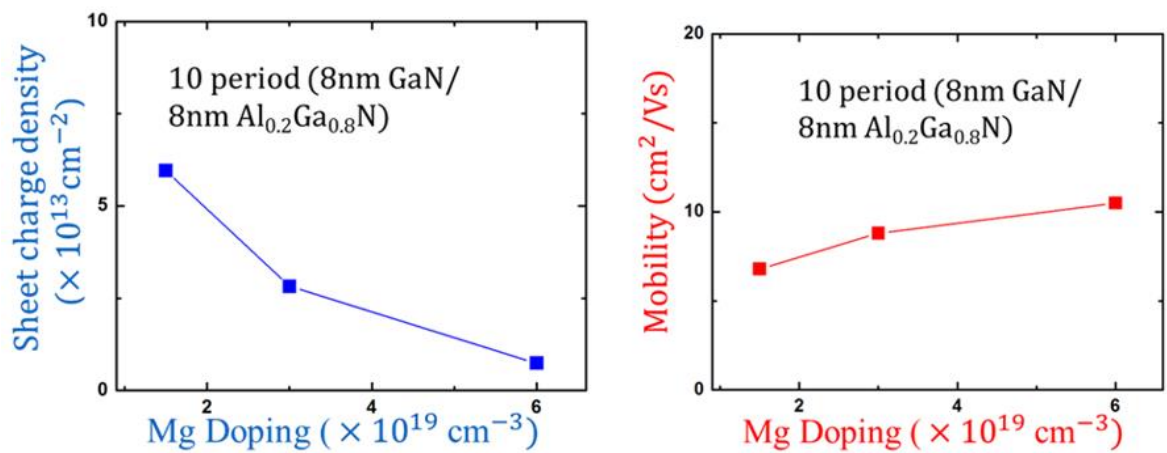

**Figure 3.** Room temperature Hall measurements yielding **(left)** total sheet charge density and **(right)** mobility of holes as a function of Mg doping in the sample

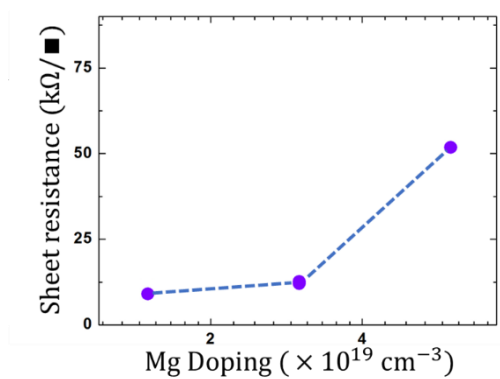

**Figure 4.** Sheet resistance as a function of Mg doping derived from the TLM measurements

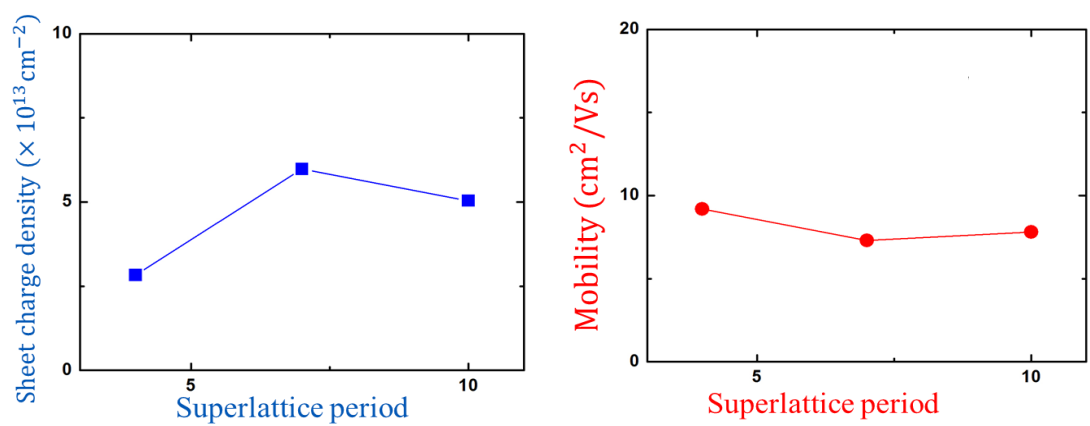

**Figure 5.** Room temperature Hall measurements yielding **(left)** total sheet charge density and **(right)** mobility of holes as a function of p-(GaN/AlGaN) SL periods

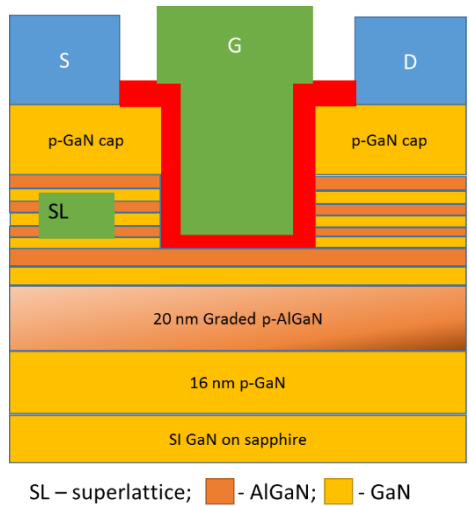

**Figure 6.** 3-period SL gate recess device design used for pFET fabrication.

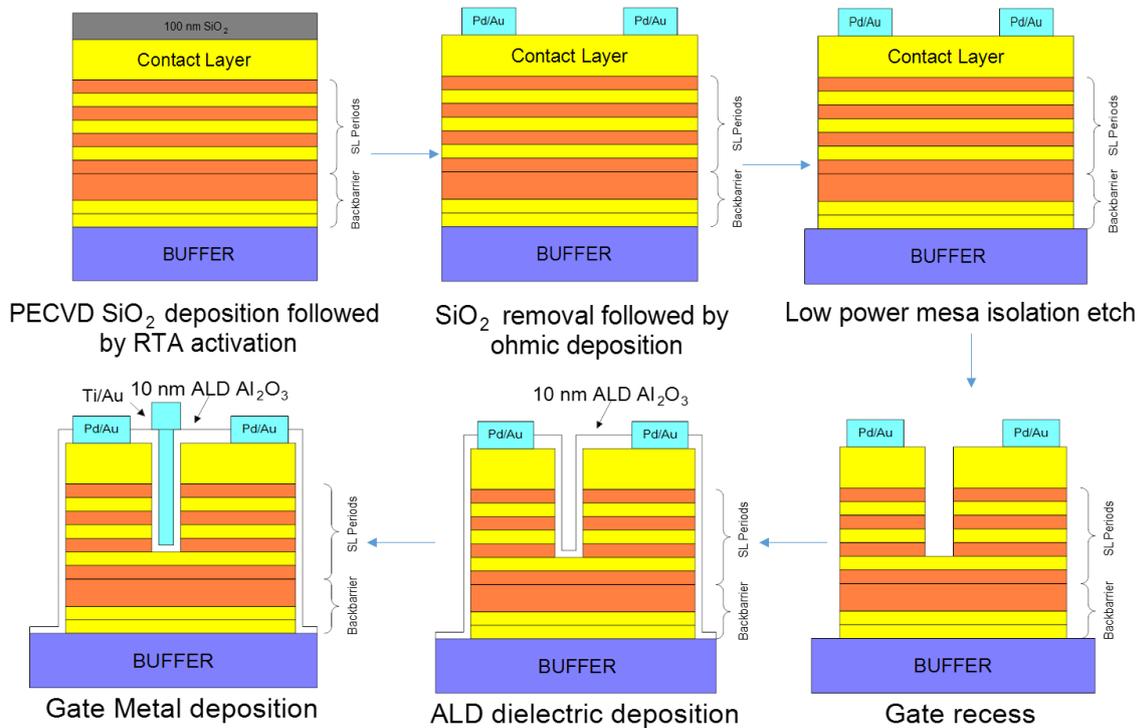

**Figure 7.** Device fabrication process flow for the p-(GaN/AlGaN) SL FET without TMAH treatment.

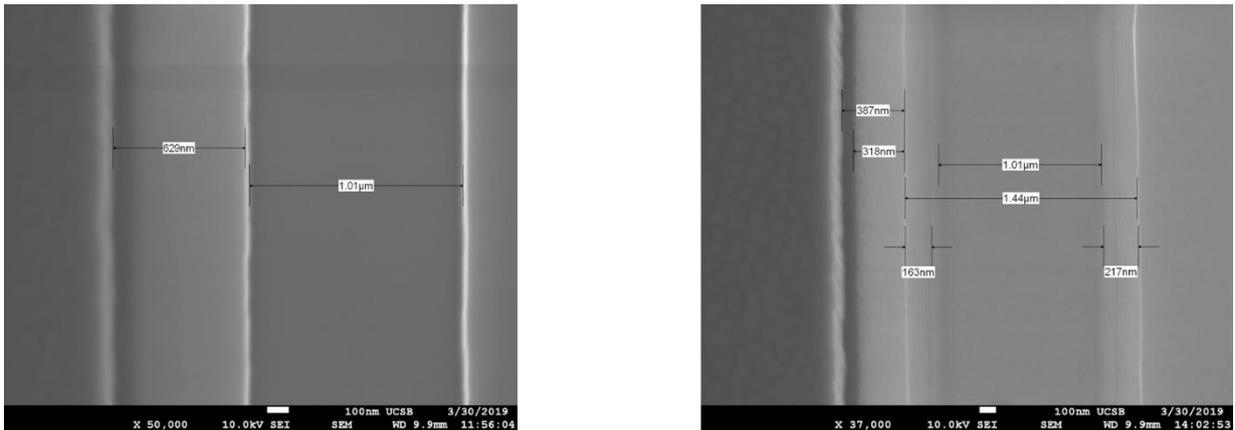

**Figure 8.** SEM images (left) before and (right) after TMAH treatment at $80^0$C for 30mins

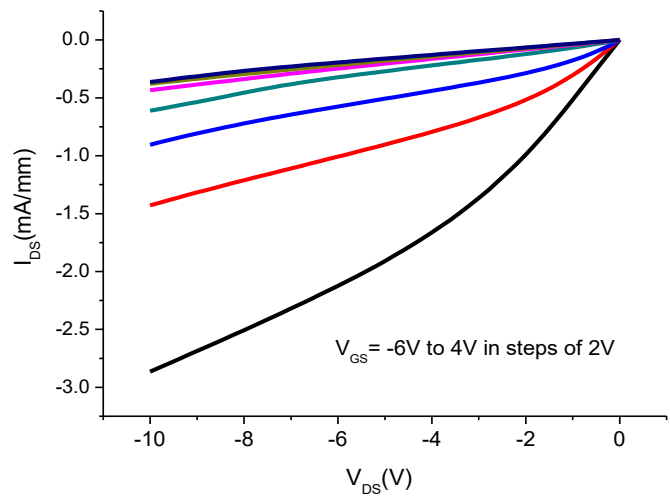

**Figure 9.** Output characteristics of the pFET fabricated without TMAH treatment

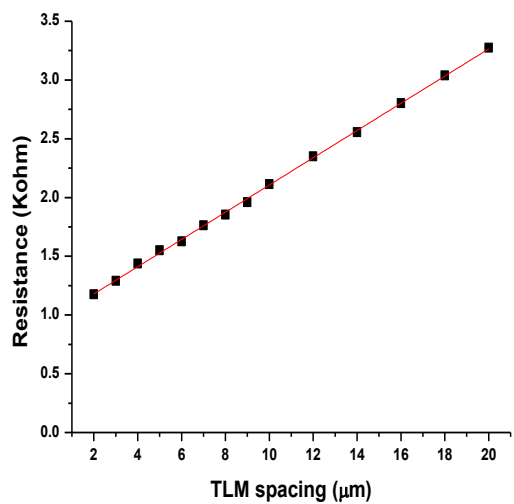

**Figure 10.** Resistance as a function of TLM spacing.

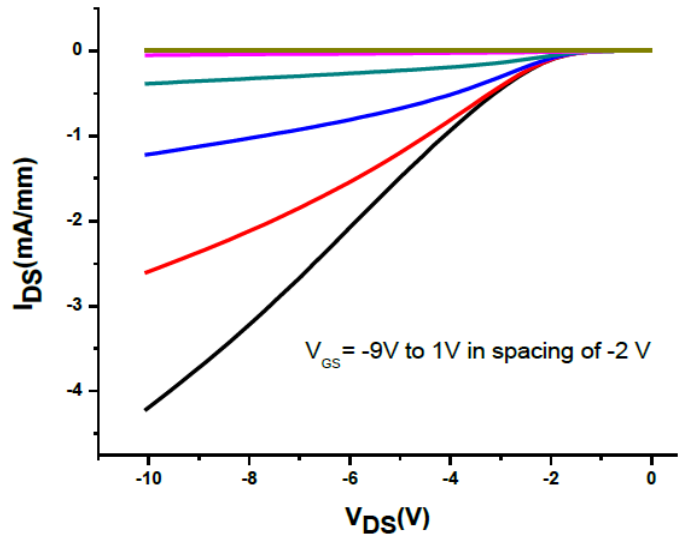

**Figure 11.** Output characteristics of the pFET fabricated with TMAH treatment

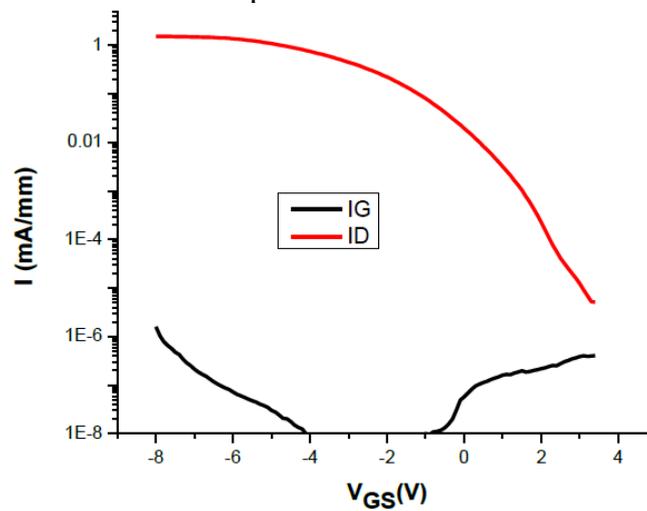

**Figure 12.** Log scale transfer characteristics of the pFET fabricated with TMAH treatment, showing five orders of current modulation and a gate leakage current orders of magnitude less than the drain current

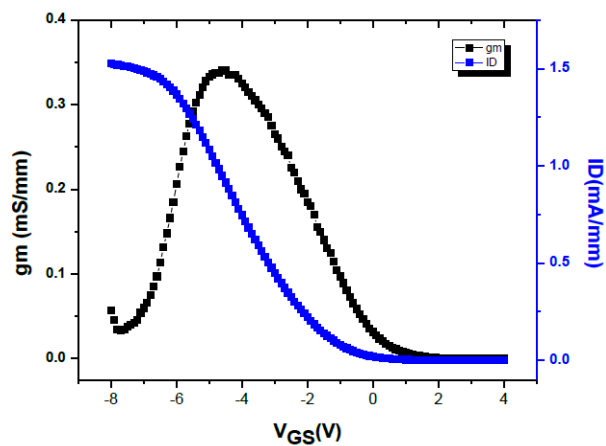

**Figure 13.** Linear scale transfer characteristics of the pFET fabricated with TMAH treatment, showing a pinch-off voltage of 2V and $g_{max}$ of 0.35 mS/mm

This paper shows the first demonstration of a p-type field effect transistor (pFET) based on an AlGaN/GaN superlattice (SL). MOCVD growth technique was used magnesium was used as the p-type dopant. With Tetramethylammonium hydroxide (TMAH) treatment during fabrication, a maximum $I_{DS}$ of 4.5mA/mm, $R_{ON}$ of 2.2kΩ.mm, and five orders of current modulation was demonstrated.

**Keyword** p-type GaN

A. Raj[1], N. Hatui[1], O. Koksaldi[1], R. Jang[1], S. Keller[1], U. K. Mishra[1], A. Krishna[1,*]

**Title**
AlGaN /GaN superlattice based p-channel field effect transistor (pFET) with TMAH treatment

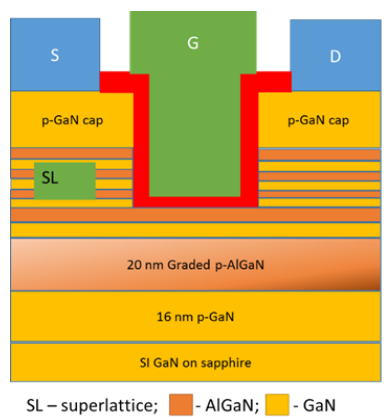